# Biologists meet statisticians: A workshop for young scientists to foster interdisciplinary team work

Benjamin Hofner[*1], Lea Vaas[*2], John-Philip Lawo[3], Tina Müller[4], Johannes Sikorski[2], and Dirk Repsilber[#5]

[*]: contributed equally
[#]: correspondence to: repsilber@fbn-dummerstorf.de, +49 38208 68 916

[1] Institut für Medizininformatik, Biometrie und Epidemiologie, Friedrich-Alexander-Universität Erlangen-Nürnberg, Waldstr. 6, 91054 Erlangen, Germany, Benjamin.Hofner@imbe.med.uni-erlangen.de
[2] Leibniz Institut DSMZ – Deutsche Sammlung von Mikroorganismen und Zellkulturen GmbH, Inhoffenstr. 7B, 38124 Braunschweig, Germany, Lea.Vaas@dsmz.de, Johannes.Sikorski@dsmz.de
[3] Global Drug Discovery Statistics, Bayer Pharma AG, 13342 Berlin, Germany, tina.mueller@bayer.com
[4] Biostatistics / CRD Global Clinical Services, CSL Behring, P.O. Box 1230, 35002 Marburg, Germany, John-Philip.Lawo@clsbehring.com
[5] Leibniz Institute for Farm Animal Biology, Genetics and Biometry, Wilhelm-Stahl-Allee 2, 18196 Dummerstorf, Germany

## Abstract

Life science and statistics have necessarily become essential partners. The need to plan complex, structured experiments, involving elaborated designs, and the need to analyse datasets in the era of systems biology and high throughput technologies has to build upon professional statistical expertise. On the other hand, conducting such analyses and also developing improved or new methods, also for novel kinds of data, has to build upon solid biological understanding and practise. However, the meeting of scientists of both fields is often hampered by a variety of communicative hurdles – which are based on field-specific working languages and cultural differences.

As a step towards a better mutual understanding, we developed a workshop concept bringing together young experimental biologists and statisticians, to work as pairs and learn to value each others competences and practise interdisciplinary communication in a casual atmosphere. The first implementation of our concept was a cooperation of the German Region of the International Biometrical Society and the Leibnitz Institute DSMZ-German Collection of Microorganisms and Cell Cultures (short: DSMZ), Braunschweig, Germany. We collected feedback in form of three questionnaires, oral comments, and gathered experiences for the improvement of this concept. The long-term challenge for both disciplines is the establishment of systematic schedules and strategic partnerships which use the proposed workshop concept to foster mutual understanding, to seed the necessary interdisciplinary cooperation network, and to start training the indispensable communication skills at the earliest possible phase of education.

**Keywords:** Interdisciplinary co-operation, biologist, statistician, mutual acceptance, communication between disciplines

# Motivation / Introduction

High-quality research requires a suitable experimental design, high-quality data, appropriate data analysis methods, a professional implementation of this analysis, and finally a sound interpretation of the results. In the field of life sciences, high-quality data stem from the experimental work, e.g., biology. Although the application of statistical methods for experimental design and data analysis has a long tradition in life sciences (Russell, 1949; Wright, 1931), this area gets the short end of the stick already in education at the university (see e.g. National Research Council, 2003; Bialek and Botstein, 2004). Thus, biologists perceive their work as mainly the realization of ideas by performance of wet-lab work. In this world, the analysis of quantitative outcomes, namely the data, is a subsequent – seemingly independent – step, which is often stepmotherly treated or even outsourced.

Suitable tools for appropriate data analysis are developed and used mainly by statisticians. Appropriate data analysis of wet life science data usually requires solid know-how in the field of statistics. Thus, the first contact between biologists and statisticians usually comes to pass in the context of consultancy (O'Day, 2001).
Biologists usually prefer to complete the laboratory work without having to think about suitable experimental designs from a statistics perspective. This perspective would involve the number of observations needed to test a hypothesis, the difference of technical and biological replicates, suitable accounting for confounding variables, multiplicity of testing or statistical analyses strategies adequate for certain experimental questions. Biologists often are not aware that statistical considerations might strongly impact their research plan. Thus, it is very common that, when a biologist contacts a statistician the first time, he or she brings a collection of already measured data, numerical results and graphics, often without an approach of possible analysis. Sometimes those structures of the experimental design which would have been important to choose the appropriate statistical analysis are only insufficiently documented. Also often, clear questions or hypotheses regarding the possible outcomes of experiments need to be formulated together with the statistician – while facing deadlines for submission of the thesis or a publication. Often biologists are not aware of potential statistical pitfalls and have no suitable basic mathematical or statistical vocabulary. This makes it hard to formulate the biological problem, while focusing on the statistical issues of design and analysis.

On the other hand, most statisticians have never spent time in a research laboratory attending experimental work. They usually do not have basic knowledge about common biological, analytical methods or about a biologist's vocabulary. Further, they have no or only a vague idea about lab techniques and procedures, typical data features such as variances or error sources, the effort of lab work or costs of experimental research projects. Thus, they hardly understand reasons for missing data, small sample sizes and the need for indirect assessments.

Consequently, it is no surprise that often a meeting of these disciplines – typically during statistical consulting – is affected by pressure of time and a mutual lack of understanding. However, it has become clear that the development of novel statistical methods strongly benefits from real data input from the life sciences as well as from basic biological knowledge on the statistician's site (Koyutürk, 2010). Regarding systems biology modelling, appropriate approaches for modelling are not within reach until bio-mathematical and molecular biological know-how is tightly combined. For a review on challenges in cross-disciplinary research in general we refer to Grigg and Pinkney (1999). For our concrete case of biologists and statisticians, apart from the reasons mentioned above, the failure of those interdisciplinary contacts might have additional other, less apparent, causes – residing merely as prejudices and inner attitudes, for each of biologists and statisticians: Biologists often do not consider data analysis as a substantial part of their work – even if it should actually be one of the first topics to think about when planning an investigation. Only seldom statistical assistance is sought. Yet, this would improve the quality of experimental results and the gain of relevant knowledge. Far too often, biologists consider statistics merely as a necessary evil on their way to publication.

Correspondingly, a typical experience of statisticians is that biologists only attend consultancy if they have to (for example due to specific referee comments) and that biologists are not really interested in sound or even sophisticated statistical analysis gaining interesting results. The perception from the statistician's point of view is that biologists only want to have a certain, pre-defined result of group differences, or "significances", i.e., significant results. Even worse, many hypotheses are not well defined – at least from the statistician's perspective, which itself is often limited in understanding biological concerns. In this case, the statistical analyses seemingly have to result in "fishing for significances" (Jelizarow *et al*., 2010). In addition, biologists seldom are able to report analysis methods and statistical parts of the results sufficiently for reproducibility. Thus, for a typical consultation session, from the consultant's point of view, a large amount of time is spent to ascertain key information from the biologist. In this situation, statisticians regularly feel exploited as only very little value is attached to their knowledge and skills from the biologist's side. They might therefore even be discouraged to attempt to understand the underlying biological question and the way of reasoning. It is partly because of such reasons that many statisticians are detracted in realization of possible benefits and potentials, which such co-operations with biologists could provide for their own work.

From the biologist's point of view, the statistician's position in a typical consulting meeting would gain a lot of trustworthiness if completed with wet-lab methodological knowledge and an understanding for the biological questions and objectives. For the biologist, the statistician, who did not investigate anything to measure the data of interest, and does not understand why just these data could be measured, has not the right to decide which data were eligible for analysis, and which data should not be used any further (or what is missing).

Only the access to and understanding of real-world data sets can keep methodical work of the statistician in good stead. Even more important, insights into biological research projects can be highly inspiring. They can result in new or improved data evaluation approaches by considering the data generating process as well as details of the biologist's interests or structures of *a priori* knowledge on the underlying biological problem.

Hence, for both, biologists as well as statisticians, the work invested for inventing new statistical methods that are adapted towards the applicability for real data objectives, pays back in multiple ways – not least in higher impact publications. However, this is only possible in tight co-operation of statisticians and experimentalists from the beginning.

All this requires a mutual and respectful team play of both disciplines. Here, however, the problems usually start. Unfortunately, at the end of many typical consulting meetings both participants might be frustrated and the mutual negative conception and prejudices are tightened. Hence, there is an urgent and strong need to develop a mutual understanding on the importance and also pitfalls of both disciplines, and to establish a working environment in which this can be achieved.

To break this vicious circle early in scientific education, we tested a concept which sets up a working situation where both disciplines have the possibility to perceive the counterpart as equal and valuable. As addressed above, the typical "consultation" is not a level-to-level meeting of experimentalist and statistician: while the experimentalist has to explain intention of the experiments, the data and the observed outcome, the statistician has to accept the data without the possibility to anticipate statistical pitfalls of the study. However, the statistician may cause large disappointment at the biologists side, if proposing not to use parts of the dataset, since the biologist looses invested time, money and personal energy (in contrast to the statistician who looses nothing).This imbalance changes only if both are equal partners in a co-operation. Such co-operation, in turn, requires readily established relationships, based on mutual trust, prior to planning any experiments.

To find a setting for promoting such a respectful co-operation, we organised a workshop for 20 biologists and 20 statisticians. Voluntary biologists from the DSMZ were recruited and statisticians (graduate or PhD students) from all over Germany were invited to apply for the workshop. Prior to the workshop, the biologists received an in-house teaching from one of the organisers to obtain basic skills in statistics and usage of the free statistical software R (R Development Core Team, 2011). All statistical analysis, both during the workshop were performed using R (version 2.14.2). During the workshop, fixed pairs of biologists and statisticians, randomly assigned to each other *a priori* by the workshop organisers, worked together for the whole time. Also, the seating arrangements in the lecture room were predefined with alternating biologists and statisticians, in order to bring forward cross-disciplinary exchange and to avoid the building of groups within the disciplines.

To evaluate the actual effects of the workshop, a survey with three questionnaires – one at the beginning, a second one at the end of the workshop, and a third one six weeks after the workshop – was conducted.

In the following, we present a summary of the workshop's concept and content of teaching. The survey's results will be discussed together with the experiences and impressions of both the organising team and participants.

## Workshop: Concept and teaching content

The key educational target for our workshop was that both partners (biologist and statistician) had to develop a strategy to be able to communicate with each other in order to manage the tasks of the programme satisfyingly for themselves. We explicitly did not aim to reach a defined educational target in terms of specific knowledge about certain topics. Secondly, the workshop aimed at raising the appreciation of the participants of the respective other scientific field. We wanted to foster a positive attitude towards learning from each other, and to build a solid basis for future co-operations. During the workshop "co-operations" in two-party teams consisting of one statistician and one biologist, the participants should perceive their partner's knowledge as different, but also as necessary and valuable to solve the common objective.

Regarding the content of the workshop, we focussed on the following objectives:

- *sensitise the participants for different aspects of trans-disciplinary communication:* language, as well as thinking ("philosophy") of the other discipline
- *teaching of basic statistics:* basic statistical testing, multiple linear regression and ANOVA, multiple testing
- *teaching of basic biology:* biological methods and problems (including PCR, phenotype microarrays, cryo-conservation of plant cells, oncogenes, phylogenetic trees, morphological heterogeneity in bacterial cells), and a day in the wet-lab conducting experiments together
- *lecture example of statistical bioinformatics as a field of research where biology and statistics competences are closely melted together:* analysis of microarray gene expression experiments
- *social events* in the evening (meeting in a casual environment).

In total, the workshop comprised four days. The detailed programme is given in Appendix A. The choice of contents for the lectures/labs should reflect both basic statistics for the biologists (and exercises, where the statisticians could teach "their" biologists), as well as biology for the statisticians (where the biologist took over the teacher's role, enabling basic understanding of molecular biology and lab practise for "his" or "her" statistician). In addition, during the lectures on statistical bioinformatics for microarray analysis, both partners of the team could practise their interaction during working out exemplary R-scripts analysing an example microarray dataset.

We chose the free statistical software R (R Development Core Team, 2011) for all applications of statistical analysis during the workshop, as nowadays it is the de facto standard in many statistical fields and – as it is freely available – it was accessible for the participants without extra cost. Since add-on packages for nearly every possible analysis are available, the participants find a powerful and reliable tool which enables them to start their own data analysis attempts, irrespectively from the specific experimental topic. Thus, learning processes as well as work in progress, like our workshop, are optimally supported by the software. As a consequence, biologists are enabled to find first solutions to their statistical problems and they are encouraged to think about the design of their experiments also from a statistical perspective.

## Survey results and experiences summary

Before and after the workshop we conducted a small survey including questions about knowledge in basic statistics, biology and laboratory experience – as well as about expectations and experiences, respectively, regarding the workshop. We assessed the participants' agreement with 14 items before and 13 after the workshop. Each item could be rated on a scale from 1 (full disagreement) to 5 (full agreement). All together, completed questionnaires for both time points from 17 biologists and 18 statisticians where collected. The questionnaires as well as detailed results are given in Appendix B. A third short questionnaire was sent out six weeks after the workshop to evaluate the long-time consequences of the workshop. In the follow-up questionnaire, we primarily asked for textual feedback and tried to inquire if the new interdisciplinary contact is still actively used or if the participants plan to use this contact. All data analysis and graphical representation of the results were performed using R.

According to the self-assessment in the questionnaires, the factual knowledge of the other discipline was improved (Fig. 1) as a consequence of the course and especially the interaction with the partners. At the same time, statisticians expanded their statistical knowledge, while biologists did not improve their biological knowledge.

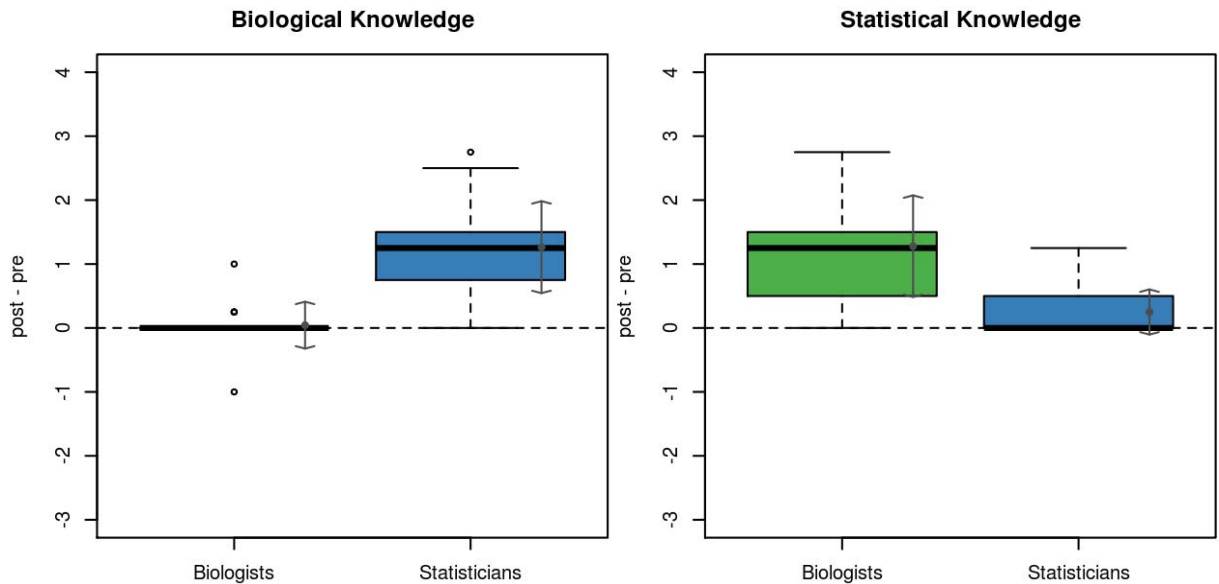

Fig.1: Boxplots and mean (±sd; in grey) of differences of questionnaire scores given by biologists/statisticians directly before and after the workshop. Higher values correspond to more knowledge.

Regarding the participants in our workshop, both, biologists and statisticians, consider statistical knowledge important for high-quality biological research. At the same time, both groups think that biological knowledge is important for the statistical analysis of biological data (Tab. 1). Hand in hand with this conception, a lack of mutual understanding exists. Both groups express a lack of ability to put oneself in the position of the other field (Tab. A.2, Appendix B). As a result of participating in the workshop, participants considered to have increased the empathy for the other field of research (Fig. 2) with the exception of one statistician whose empathy decreased during the workshop.

Tab. 1: Questionnaire scores (mean and standard deviation (SD)) evaluating the perceived importance of statistics/biology for the analysis of experimental data in biology by statisticians/biologists before the workshop. Higher values correspond to a higher agreement (min = 1, max = 5).

|   |   | Group | Mean | SD |
|---|---|---|---|---|
| Importance of... | Statistics | Biologists | 4.29 | 0.92 |
|   |   | Statisticians | 4.78 | 0.73 |
|   | Biology | Biologists | 4.29 | 0.85 |
|   |   | Statisticians | 4.28 | 1.18 |

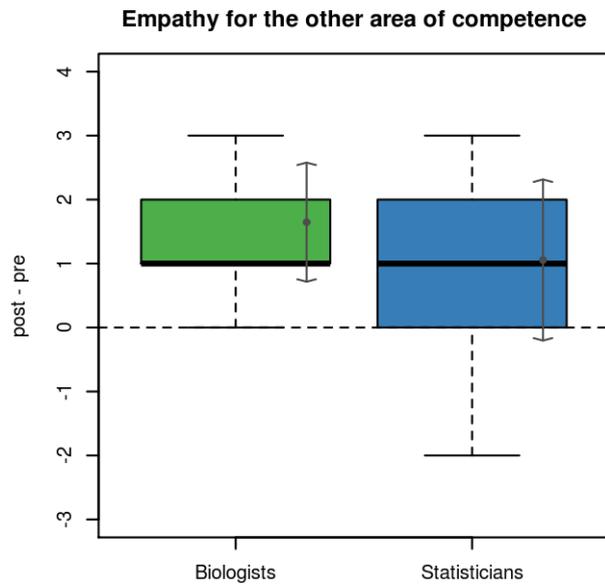

Fig. 2: Difference in questionnaire scores (boxplots and mean (±sd; in grey)) showing improvement of empathy for the respective other area of competence resulting from the workshop participation. Higher values correspond to higher empathy.

**Mutual understanding and respectful team play**

Participants from both fields expressed worries about their inability to follow the workshop *a priori*. However, after the workshop the worries were less pronounced. Most participants were well able to follow the presented contents (Fig. 3). Only two statisticians had more confidence ahead of the workshop than after the workshop. Moreover, many participants expressed their wish to participate in similar events in the future (Tab. A.2, Appendix B).

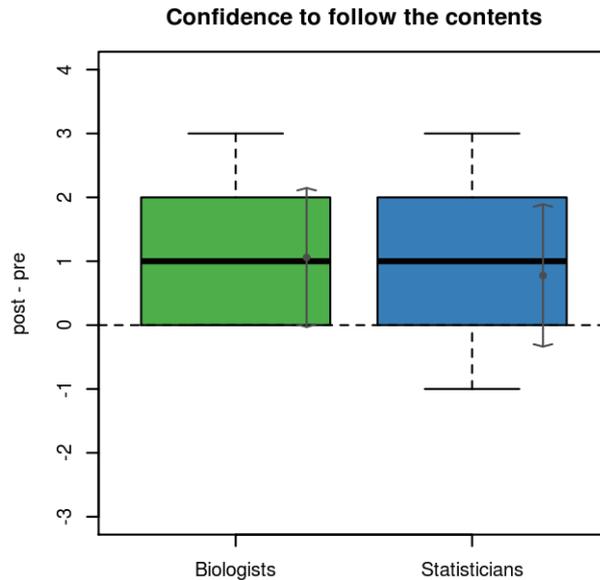

Fig. 3: Increased confidence to be able to follow the interdisciplinary contents (after – before the workshop, boxplots and mean (±sd; in grey)). Higher values correspond to greater confidence.

A further aim of this workshop was to foster long term interdisciplinary team work. The third questionnaire was answered by 21 participants. Four biologists and four statisticians (38.1 % of the respondents) stated that they kept the contact alive and are planning to do so further. Three biologists and three statisticians (28.6 %) did not get in touch so far but are planning to do so in near future. If concrete questions arise, further participants (2 biologists / 3 statisticians) are planning to use their new contacts (23.8 %). One biologist and one statistician are not intending to get in contact with the corresponding workshop partner.

## Discussion

Modern biology confronts scientists with research problems of enormous complexity comprising large data quantities from various sources (Moore and Williams 2005). To be able to push science forward, curiosity and a disposition to engage in unfamiliar areas of expertise are required. For molecular biologists, one of those "foreign fields" is the statistical analysis of experimental data, but also informatics, physics or chemistry could be referred to (O'Day et al. 2007). In fact, the importance of interdisciplinary co-operation has been emphasised throughout many disciplines consequently calling for interdisciplinary teaching at the university (e.g. Golde and Dore, 2001; Chiang, 2003; Bialek and Botstein, 2004). Not until the recent years curricula at universities began to include statistics teaching for biologists on a regular, even interdisciplinary basis (e.g. see National Research Council, 2003).

Considering the German academic landscape, at least in medical schools, the curricula are going to be reformed (Stalmeijer et al. 2007). However, at most other departments, education is still organised according to the disciplinary structure into which science and knowledge is compartmentalised (Norton 2003). Thus, especially students and young scientists are well trained only in their main subject, but to be able to cope with modern research projects they need to develop communication skills and interdisciplinary working experiences.

**Workshop concept and contents**

We developed and implemented a workshop that actually provided a framework for practising interdisciplinary team work, learningand communication, as well as to initiate a transformation for the mutual reception and understanding of biologists and statisticians. As we targeted PhD students and younger PostDocs from both disciplines, our hope was to come into action early enough to enable a sustainable basis for co-operations between experimentalists and data analysts, at an early timepoint in their career as future interdisciplinary working scientists.

In contrast to the usually tense situation of consultation, where conflicts of interest often occur and, thus, involved persons often are not working together as a team towards a common goal, we established a situation where no participant depends unilaterally on the merits of the other. The whole workshop did not aim at the completion of certain exercises or of data evaluation work flows, but was intended to build a starting point for communication on an equal level and intensive bi-directional information exchange. Therefore, we tried to arrange the workshop as an open outcome procedure disburdened from pressure of time, objectives that have to be fulfilled and mutual expectations. In this liberate working environment the participants had the chance to experience each other as experts in their field. There was room for the development of fundamental strategies for the interdisciplinary communication - namely to obtain and communicate the relevant information - which can be practised and improved iteratively. It has been observed by others that a "ready and open attitude" is a necessary precondition enabling true interdisciplinary cooperation (Wall and Shankar, 2008). Each side has to realise that the next logical step to move forward in accomplishing the desired analysis needs to cross the border to the respective other field (Hagoel and Kalekin-Fishman, 2002). Specifically for our workshop, the goal was to achieve the mutual acceptance of competences and the ability to communicate the important parts of a biological experiment, its results and the subsequent analysis.

**Survey results and experiences summary**

Our observations during the workshop, participants' statements, and the survey results indicate that we may have been successful in creating an important key experience for initiating a common basis and mutual understanding for both participating biologists and statisticians – hopefully fostering interdisciplinary networking activities for their scientific future.

Interdisciplinary co-operations can only succeed when the partners join up motivated by their own willingness and courage (Krapp et al. 1992). Once attending in interdisciplinary frameworks, the basic skill, the ability to communicate, has been shown to be a matter of training (Landis et al. 2004). Our approach sustains the hope that key parts of this communication can successfully be trained at the early stage of advanced master students, PhD students and young PostDocs, preparing for the serious "real-life" situation. The absence of pressure to deliver results and the team work, which could be considered as an opportunity rather than a pressure, contributed to a relaxed working atmosphere encouraging the participants to approach the joint venture of interdisciplinarity.

**Mutual understanding and respectful team play**

Summarizing, it has to be noted that mutual engagement is the core competence facilitating the establishment of sustainable working relationships, as in real life. It is conceivable, that even during study times, e.g. as part of a master project, true interdisciplinary cooperation should be a part of higher education approaches. Having a first own project, for which an analysis plan, the data analysis and the interpretation of results is required seems to constitute a necessary precondition to motivate students to approach the adventure of interdisciplinary communication (Wall and Shankar, 2008).

**Benefits and constraints of our approach**

Basically, with our concept we have been able to establish an atmosphere in which both statisticians and biologists were willing to resolve the distance of the both disciplines, who were courageous to explore the other discipline, and who were motivated to start a joint venture with unknown partners and unknown outcome (Barab and Landa 1997). Certainly, the age (and therefore perhaps the observed high level of flexibility) of the scientists participating in our workshop has to be seen as the result of a special selection, rather than being representative for the typical biologist or statistician. Hence, when interpreting our workshop results, we have to be aware of a – possibly severe – selection bias. Also, when trying to evaluate the chances of realising elements of our workshop for general study plans, or for planning fostering interdisciplinary work at larger institutions, one has to take into account, that, regarding the specific setting of our workshop, all participants were participating on a voluntary basis – which certainly biases their basic attitude towards the objectives of our workshop.

The survey confirmed that during the workshop both groups were able to extend their factual knowledge of the other discipline (Fig. 1). The survey surprisingly exhibited that even the statisticians extended their statistical knowledge in the course of the workshop. Here, an issue to consider might be that the statisticians had the younger average age and were predominantly students in master courses and that the focus of the lectures predominantly was on statistical issues.

Although the participants stated to appreciate the other discipline, they were timid of not being able to follow the lectures and issues of the course (Fig. 3). Both groups, statisticians and biologists, acknowledged the importance of the other field for fruitful bio-statistical research cooperations (Tab. 1). At the same time, the attendees declared a lack of ability to put themselves into the position of the other discipline. During the workshop, it became evident that this fact turned out to be one of the main impulses to attend. We need to remain aware that possibly a majority of their colleagues would not care about engaging to understand the other discipline – in spite of their own need to master interdisciplinary co-operations. Basic mathematics and statistics have become part of most biology studies at German universities since decades – on the other hand, true interdisciplinary team working remains a topic which is not yet part of advanced teaching. It remains an adventure – on both levels of personality and scientific grounds (see e.g. Wall and Shankar, 2008).

During the workshop it turned out that the scheduled time for interaction was too short. Because the content of the workshop was intended to comprise a basic set of topics of both disciplines, the daily schedule was arranged tightly, while the evening events were supposed to enable further contacts in a more casual atmosphere.

However, once become acquainted with the team partner, brisk discussions and vivid lateral information transfer started also in larger groups. Thus, the coffee and lunch breaks were extensively used to learn more about the foreign field of activity and to address numerous arising questions. Also the atmosphere during the lectures went to a more casual conversation, like the interposed question "… how do you say this in 'maths'-language?" a biologist asked during his lecture, shows. Such reactions proved our concept to work well. To improve this, in future workshops, one should allow for more free time, i.e., more breaks and time for discussions.

After the workshop, the participants of both disciplines assessed the skill of being able to put them into the other's position as improved (Fig. 2). The evening events enabled further contacts between both disciplines. Since good and respectful team play often goes beyond dealing with the factual issues, it may also involve other parts of the personality of the participating persons (see also Wall and Shankar, 2008). In general, the workshop was able to help overcoming possibly pre-existing reservations, at least partly, and, did not give rise to new ones.

For certain, effective approaches towards each other remain a question of prolonged practise for both biologists and statisticians – a workshop as ours can only be a primer to start such a process on the level of young scientists. On the level of post-graduate studies, it has already been called for extra time to accomplish the interdisciplinary challenge by Grigg et al. (2003), and this is certainly also true on later stages of the development as a scientist.

## Conclusions and Outlook

Interdisciplinary team work, involving biologists and statisticians, needs to build upon a basis of mutual understanding, both regarding the actual language of communication as well as regarding a mutual valuing for each others scientific perspectives, competences and practises. Our workshop approach targeted young scientists at the stage of early development of their careers and scientific personality where they begin to establish their scientific network, coining much of their future research activities. At this stage, the practical relevance of establishing interdisciplinary co-operations for the first time is at a maximum, as partners are not established yet, and, at the same time, urgent necessity is given.

For future workshops of this kind we propose changes to our concept at two points. First, the next workshop could focus on a more specific topic of data analysis. By concentrating on a more narrow set of biological questions and statistical analysis methods the lectures could cover more specific details and at the same time one could spend more time on discussions and interdisciplinary exchange. Second, we recommend lectures prior to the workshop, as we offered them, but for both disciplines and with some more stringent exercises. The teaching material should be provided and the potential participants could have to pass small examinations to ensure that they can manage the basic statistical and biological vocabulary and analysis problems as well. Also one should pay more attention to the practical skills of software usage of all participants. To use the workshop time optimally, no time should have to be spent to explain the basics of the software, when this can easily be managed prior to the workshop. With these recommendations, we are looking forward to advertising our approach for reissue, both in our own scientific networks as well as for other institutions.

# Appendix A – Detailed workshop programme

Day 0 (only for biologists):

- lectures on statistical testing with exercises (J.-P. Lawo)
- seminar on critical evaluation of "methods" parts (with focus on design and statistical analysis) in selected publications of molecular biology (L. Vaas & J. Sikorski)

Day 1:

- lecture and sketch about communication between biologist and statistician (lecture: T. Müller, sketch: J.-P. Lawo & J. Sikorski)
- lectures on linear modelling with exercises (B. Hofner)
- lecture on multiple testing with exercises (T. Müller)
- social event (evening)

Day 2:

- "wet-lab" experiments (choice between 14 projects, prepared especially for this workshop by the biologist participants)
- biology lecture I: Polymerase Chain Reaction (PCR)
- biology lecture II: Phenotype Microarray
- biology lecture III: Phylogenetic analyses of frogs
- biology lecture IV: Identification of unknown oncogenes in T-cell leucemia
- biology lecture V: Analysis of morphological heterogeneity in bacterial cultures
- biology lecture: Cryo-conservation of plant cells
- time for discussions and analysis of experimental results

Day 3:

- discussion of experimental results and analyses
- lectures on statistical bioinformatics for microarray analysis with exercises (D. Repsilber)
- feedback discussion

# Appendix B – Questionnaires

The questionnaires are given in Tab. A.1. The participants could rate each item from full disagreement to full agreement on a scale from 1 to 5, where 5 is agreement. Items marked with $^{(*)}$ are recoded to reflect the same directions as the other items (i.e., a positive formulation). In a follow-up questionnaire we asked the participants to give us feedback on the content of the workshop and to name possible improvements. Furthermore, we asked the participants if they plan to contact their team buddy or if they already had contact after the workshop.

Items 1-8 were used to build two scores, which represents the self-assessed factual knowledge of the participants. The first four items represents biological knowledge, the second four items statistical knowledge. The scores are the averaged agreement values over the corresponding 4 items. Thus, the scores range between 1 and 5, where 1 again represents disagreement, i.e., no factual knowledge, and 5 represents full agreement with the items, i.e., complete factual knowledge. The other items were considered separately. Tab. A.2 gives an overview of the results for all participants separately for biologists and statisticians. Tab. A.3 gives the results of the follow-up questionnaire.

Tab A.1: Translated and reordered version of the questionnaires. Items 1 to 8 represent factual knowledge which was inquired before and after the workshop. Items 9 to 11 were also inquired before and after the workshop. Before the workshop item 12 and 13 represent the mutual perception and possible prejudices ahead of the workshop. After the workshop items 12 and 13 represent the perception of the workshop. Item 14 was also assessed ahead of the workshop only.

**Questionnaire 1: Ahead of the Workshop**
1. I have practical wet-lab experience
2. I have never analyzed biological data $^{(*)}$
3. I have already planed biological experiments
4. I do not know what is meant by a "biological pathway" $^{(*)}$

5. I know what a "linear model" is
6. I have already applied a statistical test myself
7. I know what is meant by the "multiple testing problem"
8. I know why it is necessary to have an appropriate sample size calculation

9. I am able to find examples for hypotheses for the field of biology/statistics (opposite profession)
10. I feel comfortable with putting myself in the position of a biologist/statistician and the way of thinking/working in this field (opposite profession)
11. I fear not being able to follow the workshop's contents $^{(*)}$

12. I think statistical knowledge is indispensable to be able to deliver good biological research results
13. I do not think that biological knowledge is so important for being able to analyze data $^{(*)}$
14. I already wrote own computer programmes

**Questionnaire 2: After the Workshop**
1. I have practical wet-lab experience
2. I have never analyzed biological data $^{(*)}$
3. I have already planed biological experiments
4. I do not know what is meant by a "biological pathway" $^{(*)}$

5. I know what a "linear model" is
6. I have already applied a statistical test myself
7. I know what is meant by the "multiple testing problem"
8. I know why it is necessary to have an appropriate sample size calculation

9. I feel comfortable with putting myself in the position of a biologist/statistician and the way of thinking/working in this field (opposite profession)
10. I am able to find examples for hypotheses for the field of biology/statistics (opposite profession)
11. I was able to follow the contents of the workshop

12. The quality of the talks / experiments / exercises was good
13. I would like to attend another workshop of this kind

Tab. A.2: Detailed results from questionnaires before (pre) and after (post) the workshop in the groups of biologists and statisticians. The table shows mean values and standard deviations (SD) separately for n = 17 biologists and n = 18 statisticians. Higher values correspond to a higher agreement (min = 1, max = 5) with the, potentially recoded, item. For the items see Tab. A.1.

|  | | Pre | | Post | | Post – Pre | |
| --- | --- | --- | --- | --- | --- | --- | --- |
| | Group | Mean | SD | Mean | SD | Mean | SD |
| Item 1 | Biologists | 4.82 | 0.53 | 4.88 | 0.33 | 0.06 | 0.24 |
| | Statisticians | 1.44 | 0.92 | 3.33 | 1.37 | 1.89 | 1.41 |
| Item 2 | Biologists | 4.94 | 0.24 | 4.94 | 0.24 | 0.00 | 0.35 |
| | Statisticians | 3.94 | 1.47 | 4.22 | 1.56 | 0.28 | 1.18 |
| Item 3 | Biologists | 4.88 | 0.33 | 4.59 | 1.00 | -0.29 | 0.77 |
| | Statisticians | 1.22 | 0.55 | 2.00 | 1.50 | 0.78 | 1.31 |
| Item 4 | Biologists | 3.88 | 1.80 | 4.29 | 1.45 | 0.41 | 1.62 |
| | Statisticians | 2.00 | 1.33 | 4.11 | 1.18 | 2.11 | 1.60 |
| Item 5 | Biologists | 3.24 | 1.35 | 4.71 | 0.47 | 1.47 | 1.42 |
| | Statisticians | 4.94 | 0.24 | 5.00 | 0.00 | 0.06 | 0.24 |
| Item 6 | Biologists | 4.18 | 1.42 | 4.71 | 0.69 | 0.53 | 0.94 |
| | Statisticians | 4.83 | 0.51 | 5.00 | 0.00 | 0.17 | 0.51 |
| Item 7 | Biologists | 2.47 | 1.23 | 4.35 | 0.70 | 1.88 | 1.41 |
| | Statisticians | 4.61 | 0.61 | 4.94 | 0.24 | 0.33 | 0.59 |
| Item 8 | Biologists | 3.29 | 1.36 | 4.53 | 0.62 | 1.24 | 1.39 |
| | Statisticians | 4.56 | 0.98 | 5.00 | 0.00 | 0.44 | 0.98 |
| Item 9 | Biologists | 2.82 | 1.33 | 4.06 | 0.66 | 1.24 | 1.20 |
| | Statisticians | 3.00 | 1.14 | 4.28 | 0.89 | 1.28 | 1.02 |
| Item 10 | Biologists | 2.41 | 1.18 | 4.06 | 0.56 | 1.65 | 0.93 |
| | Statisticians | 2.67 | 1.03 | 3.72 | 0.57 | 1.06 | 1.26 |
| Item 11 | Biologists | 2.53 | 1.07 | 3.59 | 0.62 | 1.06 | 1.09 |
| | Statisticians | 3.56 | 1.25 | 4.33 | 0.59 | 0.78 | 1.11 |
| Item 12 (only pre) | Biologists | 4.29 | 0.85 | | | | |
| | Statisticians | 4.28 | 1.18 | | | | |
| Item 13 (only pre) | Biologists | 4.29 | 0.92 | | | | |
| | Statisticians | 4.78 | 0.73 | | | | |
| Item 12 (only post) | Biologists | | | 4.76 | 0.56 | | |
| | Statisticians | | | 4.61 | 0.50 | | |
| Item 13 (only post) | Biologists | | | 4.53 | 0.72 | | |
| | Statisticians | | | 4.61 | 0.61 | | |
| Item 14 (only pre) | Biologists | 2.41 | 1.62 | | | | |
| | Statisticians | 4.61 | 0.70 | | | | |
| Biological Knowledge | Biologists | 4.63 | 0.49 | 4.68 | 0.47 | 0.04 | 0.37 |
| (Items 1 to 4) | Statisticians | 2.15 | 0.69 | 3.42 | 0.90 | 1.26 | 0.72 |
| Statistical Knowledge | Biologists | 3.29 | 0.87 | 4.57 | 0.40 | 1.28 | 0.79 |
| (Items 5 to 8) | Statisticians | 4.74 | 0.35 | 4.99 | 0.06 | 0.25 | 0.35 |

Tab. A.3: Results from the follow-up questionnaire after six weeks. The table shows absolute frequencies of the participants contact plans in subgroups of biologists and statisticians.

|  |  | Biologists | Statisticians |
|---|---|---|---|
| Contact (until now): Yes |  | 4 | 4 |
|     If yes, keep contact alive: | Yes | 4 | 4 |
|  | No | 0 | 0 |
| Contact (until now): No |  | 6 | 7 |
|     If no, contact planned: | Yes | 3 | 3 |
|  | Probably | 2 | 3 |
|  | No | 1 | 1 |